# Dirac Fermions and Possible Weak Antilocalization in LaCuSb$_2$

J. R. Chamorro[1,2], A. Topp[3,4], Y. Fang[5], M. J. Winiarski[1,2,6], C. R. Ast[3], M. Krivenkov[7], A. Varykhalov[7] B. J. Ramshaw[5], L. M. Schoop[3,4], and T. M. McQueen[1,2,8,†]

[1]Department of Chemistry, The Johns Hopkins University, Baltimore, MD 21218, USA

[2]Institute for Quantum Matter, Department of Physics and Astronomy, The Johns Hopkins University, Baltimore, MD 21218, USA

[3]Max-Planck-Institut für Festkörperforschung, Heisenbergstraβe 1, D-70569 Stuttgart, Germany

[4]Department of Chemistry, Princeton University, Princeton, NJ 08544, USA

[5]Laboratory of Atomic and Solid State Physics, Cornell University, Ithaca, NY 14853, USA

[6]Faculty of Applied Physics and Mathematics, Gdansk University of Technology, ul. Narutowicza 11/12, 80-233 Gdansk, Poland

[7]Helmholtz-Zentrum Berlin für Materialien und Energie, Elektronenspeicherring BESSY II, Albert-Einstein-Straße 15, 12489 Berlin, Germany

[8]Department of Materials Science and Engineering, The Johns Hopkins University, Baltimore, MD 21218, USA

[†]mcqueen@jhu.edu

## Abstract

Layered heavy-metal square-lattice compounds have recently emerged as potential Dirac fermion materials due to bonding within those sublattices. We report quantum transport and spectroscopic data on the layered Sb square-lattice material LaCuSb$_2$. Linearly dispersing band crossings, necessary to generate Dirac fermions, are experimentally observed in the electronic band structure observed using angle-resolved photoemission spectroscopy (ARPES), along with a quasi-two-dimensional Fermi surface. Weak antilocalization that arises from two-dimensional transport is observed in the magnetoresistance, as well as regions of linear dependence, both of which are indicative of topologically non-trivial effects. Measurements of the Shubnikov – de Haas (SdH) quantum oscillations show low effective mass electrons on the order of 0.065$m_e$, further confirming the presence of Dirac fermions in this material.

## Introduction

Topological materials have become very popular over the last decade due to the new and interesting behaviors they display, such as protected edge states, novel excitations, and other, unconventional behaviors. There are now many instantiations of topologically non-trivial matter, such as two-dimensional quantum spin Hall phases[1-2], three-dimensional topological insulators[3-5], Dirac and Weyl semimetals[6-8], and nodal line semimetals[9,10]. All of these and more have been the main scope of many theoretical and experimental studies. New materials under these classes present new and exciting possibilities for integration into technology.

Of the many topological classes, Dirac materials present particularly interesting possibilities. In these materials, the energy spectrum of low-energy electrons can be described by the relativistic Dirac equation, as opposed to the conventional Schrödinger equation. These Dirac fermions can then give rise to a variety of interesting phenomena, such as large, linear magnetoresistances[7,11], quantum Hall effects[12], and high carrier mobilities arising from topological protection[13,14].

There exist many theoretically and experimentally verified examples of Dirac materials. Aside from the Dirac fermions found on the surfaces of topological insulators, these can also be observed in two-dimensional materials such as graphene[15], or in bulk three-dimensional Dirac semimetals such as Na$_3$Bi[6] and $Cd_3As_2$[7,11,13]. Recently Dirac fermions have been found in '112' materials with two-dimensional structures, such as layered $AMnSb_2$ and $AMnBi_2$ (A = Ca, Sr, Ba, Eu, Yb)[16-24] and $LaAgSb_2$ and $LaAgBi_2$[25-27].

The study of Dirac materials often requires a series of complex measurements to establish non-trivial band topology. Their electronic band structures can be calculated using Density Functional Theory (DFT), but often require expensive functionals to ensure a high degree of accuracy. By using angle-resolved photoemission spectroscopy (ARPES), this band structure can be experimentally accessed, which can then be compared to a calculated one for complete understanding. In addition, given the small size of most Dirac Fermi surfaces, these materials display quantum oscillations at relatively low magnetic fields. ARPES and quantum oscillations measurements, in conjunction with carrier transport measurements are therefore powerful tools to identify the presence of Dirac fermions.

Here we present ARPES, transport, and quantum oscillation data on $LaCuSb_2$, a tetragonal 112 material with an Sb square lattice. We find two-dimensional weak antilocalization and linear magnetoresistance through transport measurements and confirm the presence of linearly dispersing Dirac bands in the band structure. We observe Shubnikov-de Haas oscillations in magnetic fields up to $B = 12$ T and as high as $T = 30$ K and determine a light effective mass of $m^* = 0.065 m_e$, as expected for Dirac fermions. Our results demonstrate the presence of Dirac fermions in $LaCuSb_2$.

**Results and Discussion**

$LaCuSb_2$ crystallizes in the tetragonal space group $P4/nmm$ (no. 129) and contains alternating layers of a two-dimensional layer of $CuSb_2$ tetrahedra and a two-dimensional Sb square net, separated by La atoms, as shown in **Figure 1a**. It has been reported that variations in Cu content can lead to dramatic changes in the lattice parameters in this material, but our lattice parameters of $a = 4.3828(2)$ and $c = 10.2097(7)$ are consistent[28] with a Cu occupancy of 1.0. In addition, small amounts of excess Sb can be found in all synthesized samples, and is limited to a maximum of 1.5% by mass.

It is the bonding of the Sb atoms in the square net layers that can lead to the formation of linearly dispersing Dirac or Weyl crossings in the electronic band structure due to the symmetry of the Sb $p$ band overlap[10,29,30]. The bandwidth of these linearly dispersing bands is determined by the Sb-Sb interatomic distance, which is 3.08Å in $LaCuSb_2$. This distance is shorter than in some of the other aforementioned 112 Dirac materials, and thus results in linearly dispersive bands with relatively large bandwidths[16-27]. Using a path indicated by the primitive tetragonal Brillouin zone shown in **Figure 1b**, the calculated DFT band structures for $LaCuSb_2$ excluding and including spin-orbit coupling (SOC) are shown in **Figure 1c** and **Figure 1d**, respectively. The two-dimensionality of the structure is reflected in the band structure as the bands become flat and minimally dispersive in reciprocal space directions along $k_z$, such as $\Gamma - Z$ and $M - A$. Dirac fermion behavior may arise as a result of the linearly dispersive bands that cross at certain points in the Brillouin zone, which has also been observed in some of the aforementioned square-net materials. In $LaCuSb_2$, however, there are also several parabolic bands crossing the Fermi level that are primarily Cu-$d$ bands, which may result in some conventional, multiband effects. Given that Cu is $d^{10}$ in this system, no magnetism or strong electron correlations are expected.

To further understand the electronic structure of materials, ARPES measurements were performed and compared to DFT calculations. **Figure 2a** shows a constant energy cut at the Fermi level ($E_i = 0$ eV) of the first BZ and its vicinity, elucidating the nature of the $LaCuSb_2$ Fermi surface. It consists of a diamond-like feature centered around the $\overline{\Gamma}$ point connected to the next BZ at the $\overline{X}$ points, and contains no bands around the $\overline{M}$ point. This kind of diamond-shaped Fermi surface is common for square-net based nodal line materials[30-32]. The literature theoretical value of 0.719 Å$^{-1}$ for the $\overline{\Gamma}$-$\overline{X}$ distance is in accordance with the experimental data[33]. DFT calculations of the Fermi surface, shown in **Figure 2b**, display similar structural behavior but also reveal a cylindrical internal structure inside the diamond centered around $\overline{\Gamma}$, which is weakly visible in the experimental first BZ. Since intensity modulations due to matrix element effects resulting from multiple atoms in the basis affect successive BZs differently, it is also clearly resolved with a higher intensity in neighboring BZs. **Figures 2c** and **2d** show the experimental and theoretical dispersion plots along the path shown in **Figure 2b**. Both show several bands along $\overline{\Gamma}$-$\overline{X}$, dispersing linearly over a large energy range and crossing the Fermi level (two bands with comparably low intensity are marked with two arrows). To ensure an accurate comparison with the data, one must consider the effect of SOC on the band structure, since the linearly dispersing bands arise from Sb *p* bands, which can show large avoided crossings due to the large atomic mass of Sb. It should be noted that **Figure 2d** only shows the $k_z = 0$ plane, since it provides the best agreement with the ARPES data (for a comparison with the $k_z = \pi$ plane, see the **Figure S1** in the SI).

Since LaCuSb$_2$ is an inversion-symmetric material and time-reversal symmetry is not broken, all bands in the bulk band structure are at least two-fold spin degenerate. In addition, the nonsymmorphic symmetry of space group *P*4/*nmm* forces additional band degeneracies at the high symmetry points X and M (and R and A, respectively). Along the high-symmetry lines X-M and R-A, however, the degeneracy is not protected by nonsymmorphic symmetry in the presence of SOC, although the splitting of the bands appears to be quite weak and proves difficult to resolve in the experimental data. Deviations between the experimentally observed bands and the presented DFT calculations can be explained by a small change of $k_z$ throughout the BZ and a small offset from the high-symmetry plane $k_z = 0$ for the presented constant photon energy of 80 eV. Overall, however, the experimental band dispersions are in good agreement with the DFT calculations, confirming the presence of linearly dispersing bands belonging to Dirac crossings that can give rise to Dirac fermions in LaCuSb$_2$.

Transport measurements of LaCuSb$_2$ provide insight into its electronic behavior. The resistivity of LaCuSb$_2$ at room temperature amounts to 1.3 mOhm-cm. The resistivity decreases with temperature, as expected for a metal, and generally increases with applied magnetic field, as shown in **Figure 3a**. The residual resistivity ratio (RRR) is 8.5, as given by $\rho_{250K}/\rho_{3K}$, and is indicative of a high-quality crystal. The magnetoresistance (MR) of LaCuSb$_2$, however, shows more interesting behavior. At low magnetic fields and low temperatures, the MR defined as $(\rho(H) - \rho(0))/\rho(0) \times 100$ increases dramatically before becoming linear, as shown in **Figure 3b**. With increasing temperature, the region of linearity is suppressed, as well as the magnitude of the sudden upturn in MR. This sudden upturn in MR has been observed in other systems and can be explained by quantum interference effects such as weak localization or weak antilocalization, the latter of which has often been observed in the 2D Dirac surface states of topological insulators and in 3D Dirac semimetal thin films[34-37]. Furthermore, linear magnetoresistance has been known to arise in topological systems, and its presence in LaCuSb$_2$ further gives evidence for the presence of Dirac fermions in this system. Hall measurements of LaCuSb$_2$ indicate that the dominant carriers are electrons, given the negative slope of $R_{xy}$, with a relatively high carrier concentration of $2.74 \times 10^{18}$ cm$^{-3}$ at 3 K, as shown in **Figure 3c**. This value requires a consideration of a large Hall angle given the large difference in $\rho_{xx}$ vs. $\rho_{xy}$, and complicates symmetrization of the data, resulting in apparent non-linearity. This is likely due to the presence of other bands at the Fermi level which act to enlarge the Fermi surface from that expected for a simple Dirac system. Further studies are required to investigate whether the Hall behavior is intrinsically linear or not.

As aforementioned, a large upturn is observed in the magnetoresistance of LaCuSb$_2$, especially at low temperatures. This upturn can be ascribed to weak antilocalization, a phenomenon which occurs in the quantum diffusive regime of metallic materials, whereby the electron phase coherence lengths are longer than the electron mean free path lengths. Applying a magnetic field and consequently breaking time reversal symmetry results in a dramatic decrease in conductivity with small applied fields, or a converse increase in resistivity, i.e. what is observed in our magnetoresistance data. Weak antilocalization has been observed in many topological materials harboring Dirac fermions because of the $\pi$ Berry phase picked up after circulating around the Fermi surface. This $\pi$ Berry phase then generates destructive quantum interference which can suppress backscattering and lead to an increase in conductivity with decreasing temperature. However, applying a magnetic field and breaking time-reversal symmetry negates this effect and results in a large decrease in the conductivity with small applied fields.

Based on Hikami-Larkin-Nagaoka (HLN) theory, the weak antilocalization in LaCuSb$_2$ can be understood by fitting the magnetoconductivity to the following:

$$\Delta\sigma(B) = \frac{\alpha e^2}{2\pi^2 \hbar}\left[\psi\left(\frac{1}{2} + \frac{\hbar}{4eBL_\varphi^2}\right) - \ln\left(\frac{\hbar}{4eBL_\varphi^2}\right)\right]$$

Where $\alpha = -\frac{1}{2}$ due to the $\pi$ Berry phase, $\psi$ is the Digamma function, and $L_\varphi$ is the phase coherence length. The temperature dependence of $L_\varphi$ can be seen in **Figure 3d**. It follows a $T^{-0.58(4)}$ law, which is very close to the $T^{-0.5}$ relation expected for a two-dimensional system[34]. This indicates that the weak antilocalization in this material is two-dimensional, which may be due to the dimensionality of the Sb square-lattice. However, weak antilocalization in bulk materials is uncommon, and has only been observed in a very small number of bulk materials. Weak antilocalization effects are more commonly observed in topological insulator films, due to the pronounced effect of the Dirac surface states due to the thinness of the samples[35-37]. An alternative explanation, based on a two-band model, can be found in the supplementary information.

In LaCuSb$_2$, periodic SdH quantum oscillations in the resistivity at high magnetic fields can be observed, shown in **Figure 4a** for $T$ = 0.27 K as a function of inverse field. The single crystal structure, patterned by focused-ion-beam lithography, used to measure these is shown in **Figure 4b**. These oscillations are due to the cyclotron motion of carriers on the Fermi surface. Performing a Fast Fourier Transform (FFT) operation on SdH data collected at temperatures up to $T$ = 30 K reveals the amplitudes of dominant frequencies of oscillation, as shown in **Figure 4c**. The main frequency of oscillation is observed to be 49.6 T. Lastly, by considering the temperature dependence of the 49.6 T FFT amplitude shown in **Figure 4c**, the effective carrier mass $m^*$ can be obtained by fitting to the Lifshitz-Kosevich theory, as shown in **Figure 4d**[38]. This value was found to be $m^* = 0.065 m_e$, where $m_e$ is the electron rest mass. Through the oscillation frequency, we determine the Fermi momentum $k_F$ to be equal to $3.88 \times 10^8$ m$^{-1}$. By considering the Onsager relation for the carrier concentration of $n = [2/(2\pi)^3](4\pi/3k_F^3)$, we obtain $n = 1.97 \times 10^{18}$ cm$^{-3}$, which is in agreement with the Hall measurement result. It should be noted that this applies to a isotropic three-dimensional pocket and not a two-dimensional one, and thus LaCuSb$_2$ appears to display more three-dimensional behavior than other 112 systems. The ultralow carrier mass obtained is found to be smaller than that for other 112 Dirac systems[16-27], and is generally lower than one would expect for conventional (non-Dirac) carriers. These observations provide strong evidence for the presence of Dirac fermions in LaCuSb$_2$.

Our results show strong evidence for the presence of Dirac fermions in LaCuSb$_2$. In addition to this observation, there have been potentially interesting claims of intrinsic superconductivity in this system in the literature, under $T$ = 0.9 K[39,40], that in the context of our discovery, may present an interesting example of a material at the interface of superconductivity and topology[41]. Our measurements, however, of both resistivity and specific heat in multiple samples down to temperatures far below $T = 0.9$ K, have shown no anomaly consistent with a superconducting phase transition. Subsequent studies could vary the synthesis techniques and search for superconductivity in this interesting system.

**Experimental**

Centimeter-sized single crystals of $LaCuSb_2$ were grown by the flux method using antimony[42]. Stoichiometric amounts of La (99.99%) and Cu (99.99%) were placed in an alumina crucible, and Sb (99.999%) was added in thirty-fold molar excess as both reagent and flux. The crucible was placed inside a sealed fused silica tube under vacuum, and the reaction was stepped to 1050°C, soaked for 12 hours, then cooled slowly to 650°C at a rate of 5°C per hour. The reaction vessel was then centrifuged to remove excess Sb flux. Large crystals, limited mainly in size by the crucible, can be obtained. The crystals are visibly layered and easy to cut and handle, and are air and moisture stable.

Representative crystals were ground and their crystal structure determined using X-ray diffraction on a laboratory Bruker D8 Focus diffractometer (Cu tube, $K\alpha_1$ 1.540596 Å, $K\alpha_2$ 1.544493 Å) with a LynxEye detector. The structure was found to be consistent with stoichiometric $LaCuSb_2$ based on previous structural studies of this material[28,33].

Electronic transport and heat capacity data were collected on crystals in a Quantum Design physical properties measurement system. Resistivity measurements used standard four probe geometry on a bar shaped cut crystal, whereas Hall measurements were performed using a crystal cut in the shape of a square, with leads attached in the proper Hall geometry. Angular dependent transport data was collected using the Quantum Design horizontal rotator option. Heat capacity measurements were performed under $T = 3$ K using a Quantum Design dilution refrigerator.

Electronic and band structure calculations were performed on $LaCuSb_2$ by means of DFT with the local density approximation (LDA), using the Elk all-electron full-potential linearized augmented-plane wave plus local orbitals (FP-LAPW+LO) code[43]. Calculations were performed both with and without SOC using an $8 \times 8 \times 8$ $k$-point mesh.

ARPES spectra were recorded on the UE112-PGM2a beamline at Bessy II in Berlin, Germany. The utilized $1^2$ endstation is equipped with a Scienta R8000 detector. Samples were cleaved in-situ in ultra-high vacuum (low $10^{-10}$ mbar pressure range) and measured at a temperature of 40 K.

Measurements of SdH oscillations were performed on a single-crystal structure prepared using focused ion beam lithography. Four-point resistance measurements of the c-axis resistivity were made using an excitation current of 150 μA, with magnetic fields up to 12 T applied along the c-axis, at temperatures between $T = 0.27$ and 30 K.

**Acknowledgments**

This work was supported as part of the Institute for Quantum Matter, an Energy Frontier Research Center funded by the U.S. Department of Energy, Office of Science, Office of Basic Energy Sciences, under Award DE-SC0019331. L. M. S. was supported by NSF through the Princeton Center for Complex Materials, a Materials Research Science and Engineering Center DMR-1420541, and by a MURI grant on Topological Insulators from the Army Research Office, grant number ARO W911NF-12-1-0461. A. T. was supported by the DFG; proposal no. SCHO 1730/1-1. Y. F. and B. J. R. are supported by the National Science Foundation under Grant No. 1752784. We thank HZB for the allocation of synchrotron radiation beamtime.

**Supplemental Materials**

See the supplementary materials for a more in-depth discussion of the ARPES findings and a discussion of other possible sources of linear magnetoresistance in this material.

**References**

1   Kane, C. L. & Mele, E. J. Z(2) topological order and the quantum spin Hall effect. *Phys Rev Lett* **95**, 146802 (2005).

2   Kane, C. L. & Mele, E. J. Quantum spin Hall effect in graphene. *Phys Rev Lett* **95**, 226801 (2005).

3   Hasan, M. Z. & Kane, C. L. Colloquium: Topological insulators. *Rev Mod Phys* **82**, 3045-3067 (2010).

4   Zhang, H. J. *et al.* Topological insulators in $Bi_2Se_3$, $Bi_2Te_3$ and $Sb_2Te_3$ with a single Dirac cone on the surface. *Nat Phys* **5**, 438-442 (2009).

5   Fu, L., Kane, C. L. & Mele, E. J. Topological insulators in three dimensions. *Phys Rev Lett* **98**, 106803 (2007).

6   Liu, Z. K. *et al.* Discovery of a Three-Dimensional Topological Dirac Semimetal, Na3Bi. *Science* **343**, 864-867 (2014).

7   Liang, T. *et al.* Ultrahigh mobility and giant magnetoresistance in the Dirac semimetal $Cd_3As_2$. *Nat Mater* **14**, 280-284 (2015).

8   Lv, B. Q. *et al.* Experimental Discovery of Weyl Semimetal TaAs. *Phys Rev X* **5**, 031013 (2015).

9   Fang, C., Weng, H. M., Dai, X. & Fang, Z. Topological nodal line semimetals. *Chinese Phys B* **25**, 117106 (2016).

10  Schoop, L. M., Pielnhofer, F. & Lotsch, B. V. Chemical Principles of Topological Semimetals. *Chem Mater* **30**, 3155-3176 (2018).

11  Li, H. *et al.* Negative magnetoresistance in Dirac semimetal $Cd_3As_2$. *Nat Commun* **7**, 10301 (2016).

12  Zhang, Y. B., Tan, Y. W., Stormer, H. L. & Kim, P. Experimental observation of the quantum Hall effect and Berry's phase in graphene. *Nature* **438**, 201-204 (2005).

13  Neupane, M. *et al.* Observation of a three-dimensional topological Dirac semimetal phase in high-mobility $Cd_3As_2$. *Nat Commun* **5**, 3786 (2014).

14  Bolotin, K. I. *et al.* Ultrahigh electron mobility in suspended graphene. *Solid State Commun* **146**, 351-355 (2008).

15  Novoselov, K. S. *et al.* Two-dimensional gas of massless Dirac fermions in graphene. *Nature* **438**, 197-200 (2005).

16  He, J. B. *et al.* Quasi-two-dimensional massless Dirac fermions in $CaMnSb_2$. *Phys Rev B* **95**, 045128 (2017).

17  Feng, Y. *et al.* Strong Anisotropy of Dirac Cones in $SrMnBi_2$ and $CaMnBi_2$ Revealed by Angle-Resolved Photoemission Spectroscopy. *Sci Rep-Uk* **4**, 5385 (2014).

18  Zhang, A. M. *et al.* Interplay of Dirac electrons and magnetism in $CaMnBi_2$ and $SrMnBi_2$. *Nat Commun* **7**, 13833 (2016).

19  Liu, J. Y. *et al.* A magnetic topological semimetal $Sr_{1-y}Mn_{1-z}Sb_2$ (y, z < 0.1). *Nat Mater* **16**, 905 (2017).

20  Liu, J. Y. *et al.* Nearly massless Dirac fermions hosted by Sb square net in $BaMnSb_2$. *Sci Rep-Uk* **6**, 30525 (2016).

21  Huang, S. L., Kim, J., Shelton, W. A., Plummer, E. W. & Jin, R. Y. Nontrivial Berry phase in magnetic $BaMnSb_2$ semimetal. *P Natl Acad Sci USA* **114**, 6256-6261 (2017).

22   Masuda, H. *et al.* Quantum Hall effect in a bulk antiferromagnet EuMnBi$_2$ with magnetically confined two-dimensional Dirac fermions. *Sci Adv* **2**, 1501117 (2016).

23   Wang, Y. Y., Xu, S., Sun, L. L. & Xia, T. L. Quantum oscillations and coherent interlayer transport in a new topological Dirac semimetal candidate YbMnSb$_2$. *Phys Rev Mater* **2**, 021201 (2018).

24   Liu, J. Y. *et al.* Unusual interlayer quantum transport behavior caused by the zeroth Landau level in YbMnBi$_2$. *Nat Commun* **8**, 646 (2017).

25   Wang, K. F., Graf, D. & Petrovic, C. Quasi-two-dimensional Dirac fermions and quantum magnetoresistance in LaAgBi$_2$. *Phys Rev B* **87**, 235101 (2013).

26   Wang, K. F. & Petrovic, C. Multiband effects and possible Dirac states in LaAgSb$_2$. *Phys Rev B* **86**, 155213 (2012).

27   Shi, X. *et al.* Observation of Dirac-like band dispersion in LaAgSb$_2$. *Phys Rev B* **93**, 081105 (2016).

28   Yang, X. X. *et al.* RCu$_{1+x}$Sb$_2$ (R = La, Ce, Pr, Nd, Sm, Gd, Tb, Dy, Ho and Y) phases with defect CaBe$_2$Ge$_2$ - type structure. *Mater Sci Forum* **475-479**, 861-864 (2005).

29   Tremel, W. & Hoffmann, R. Square Nets of Main Group Elements in Solid-State Materials. *J Am Chem Soc* **109**, 124-140 (1987).

30   Klemenz, S., Lei, S. & Schoop, L. M. Topological Semimetals in Square-Net Materials. *Annu. Rev. Mater. Res.* **49**(1), 185-206 (2019).

31   Schoop, L. M. *et al.* Dirac cone protected by non-symmorphic symmetry and three-dimensional Dirac line node in ZrSiS. *Nat Commun* **7**, 11696 (2016).

32   Topp, A. *et al.* Non-symmorphic band degeneracy at the Fermi level in ZrSiTe. *New J Phys* **18**, 125014 (2016).

33   Sologub, O., Hiebl, K., Rogl, P., Noel, H. & Bodak, O. On the Crystal-Structure and Magnetic-Properties of the Ternary Rare-Earth Compounds Retsb2 with Re-Equivalent-to-Rare Earth and T-Equivalent-to-Ni, Pd, Cu and Au. *J Alloy Compd* **210**, 153-157 (1994).

34   Altshuler, B. L., Aronov, A. G. & Khmelnitsky, D. E. Effects of Electron-Electron Collisions with Small Energy Transfers on Quantum Localization. *J Phys C Solid State* **15**, 7367-7386 (1982).

35   Gopal, R. K., Singh, S., Chandra, R. & Mitra, C. Weak-antilocalization and surface dominated transport in topological insulator Bi$_2$Se$_2$Te. *Aip Adv* **5**, 047111 (2015).

36   Bao, L. H. *et al.* Weak Anti-localization and Quantum Oscillations of Surface States in Topological Insulator Bi$_2$Se$_2$Te. *Sci Rep-Uk* **2**, 726 (2012).

37   Kim, Y. S. *et al.* Thickness-dependent bulk properties and weak antilocalization effect in topological insulator Bi$_2$Se$_3$. *Phys Rev B* **84**, 073109 (2011).

38   Lifshits, E. M. & Kosevich, A. M. Theory of the Shubnikov-Dehaas Effect. *J Phys Chem Solids* **4**, 1-10 (1958).

39   Lakshmi, K. V., Menon, L., Nigam, A. K., Das, A. & Malik, S. K. Magneto-resistance studies on RTSb$_2$ compounds (R=La, Ce and T=Ni, Cu). *Physica B* **223-24**, 289-291 (1996).

40   Gamayunova, N. V. *et al.* Electron-Phonon Interaction in Ternary Rare-Earth Copper Antimonides LaCuSb$_2$ and La(Cu$_{0.8}$Ag$_{0.2}$)Sb$_2$ probed by Yanson Point-Contact Spectroscopy. *Proc Int Conf Nanoma* (2017).

41   Ruszala, P., Winiarski, M. J. & Samsel-Czekala, M. Dirac-like band structure of LaTESb$_2$ (TE = Ni, Cu, and Pd) superconductors by DFT calculations. *Comp Mater Sci* **154**, 106-110 (2018).

42   Canfield, P. C. & Fisk, Z. Growth of Single-Crystals from Metallic Fluxes. *Philos Mag B* **65**, 1117-1123 (1992).

43  Mostofi, A. A. *et al.* wannier90: A tool for obtaining maximally-localised Wannier functions. *Comput Phys Commun* **178**, 685-699 (2008).

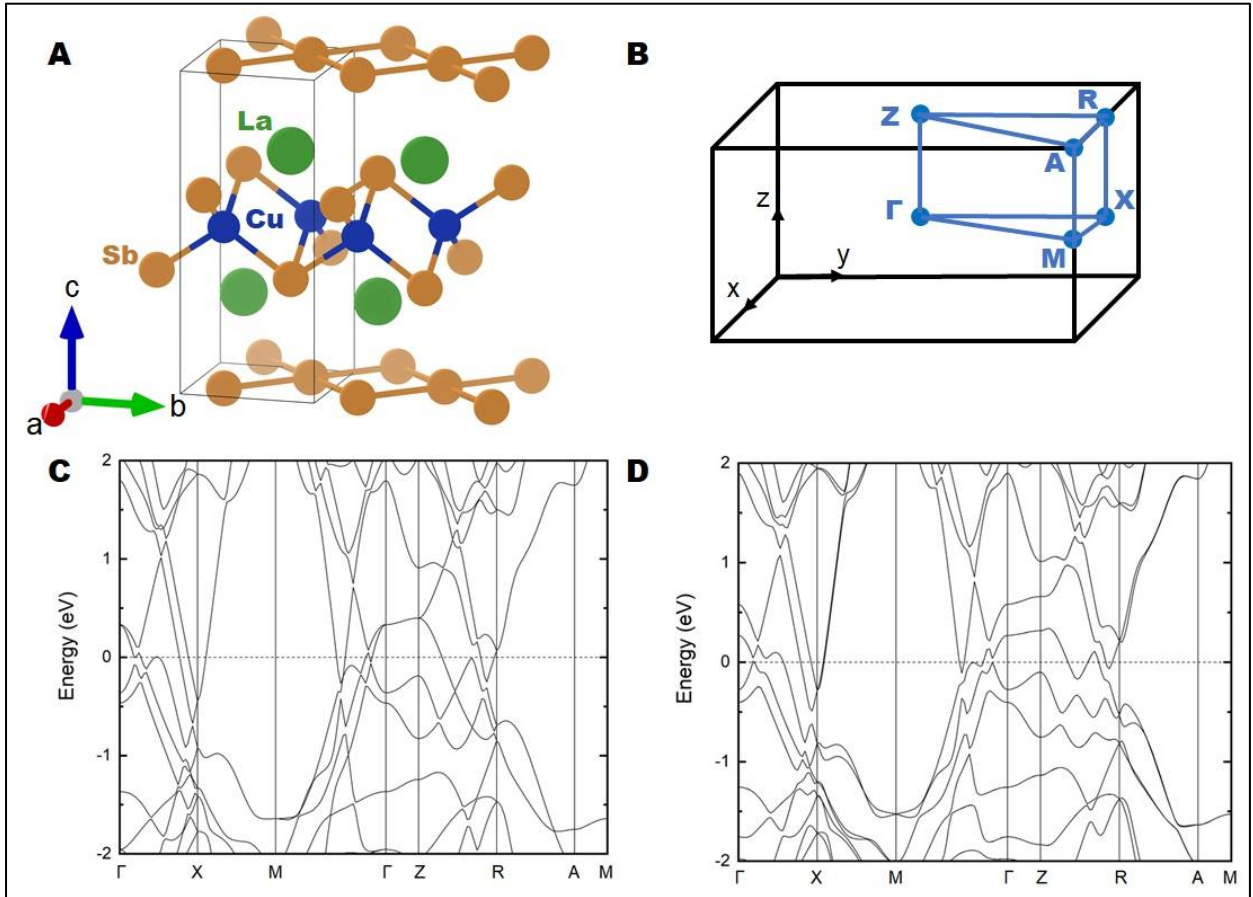

**Figure 1**. **A**. The tetragonal crystal structure of LaCuSb$_2$, showing the CuSb$_4$ layer and the Sb square-lattice layer. **B**. The first Brillouin zone for a primitive tetragonal cell, with special positions highlighted. **C** and **D** show the electronic band structure for LaCuSb$_2$ with and without spin orbit coupling, respectively. Linear bands can be observed especially along X-M and M-Γ, with bandwidths on the other of ~1 eV.

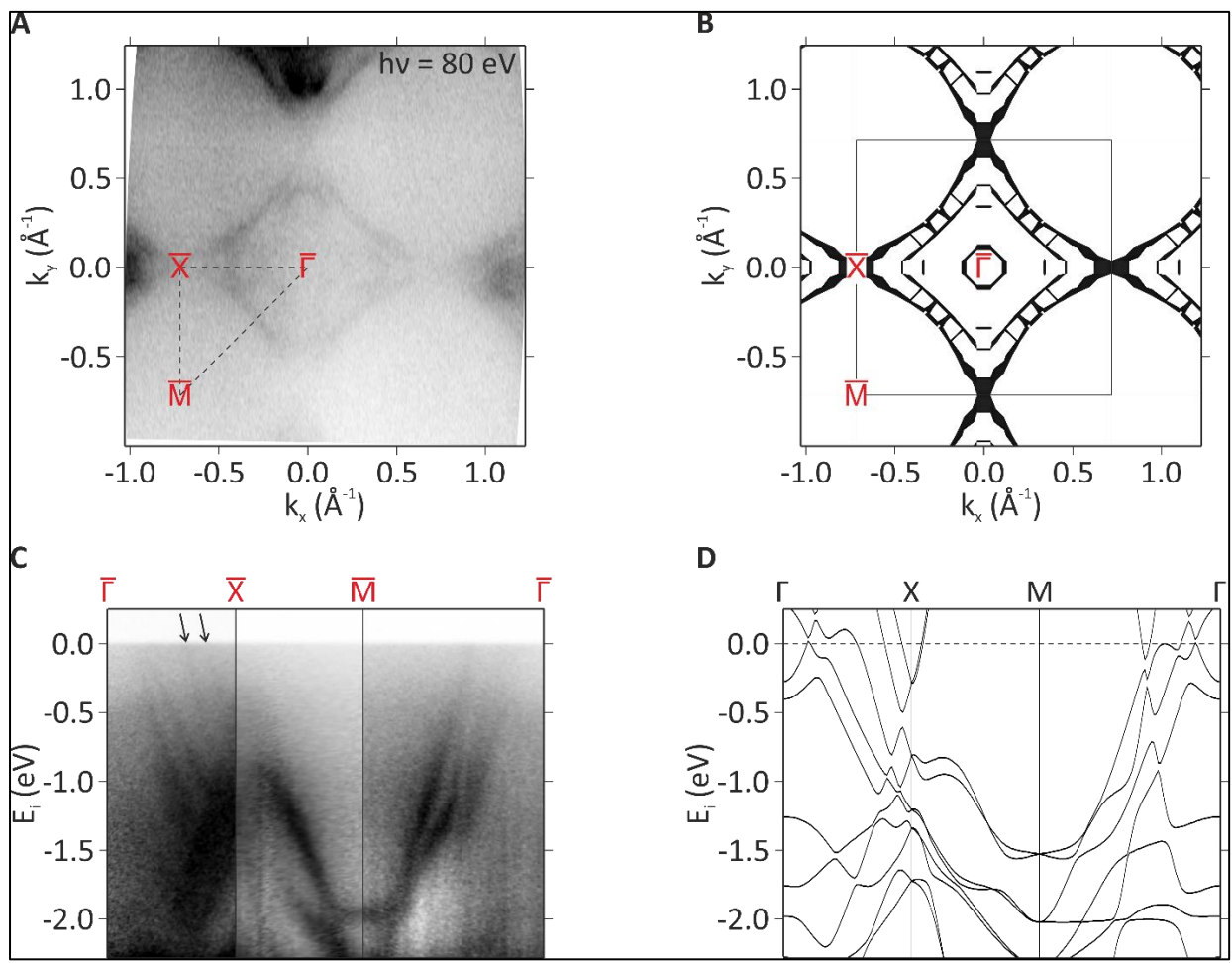

**Figure 2**. **A**. The experimentally observed Fermi surface measured at $hv$ = 80 eV, demonstrating a diamond-like shape around the $\bar{\Gamma}$ point. **B**. The calculated Fermi surface, which agrees with the experimental one in **A**. Internal features are very weak in the first Brillouin zone but are observed in the experimental second Brillouin zone, possibly due to various matrix element effects. **C**. The experimentally observed band structure for the $\bar{\Gamma}$-$\bar{X}$-$\bar{M}$-$\bar{\Gamma}$ path along the Brillouin zone. **D**. The calculated band structure for the same path as in **C**. Deviations observed in the experimental data are likely due to the presence of a small $k_z$ dispersion component. See SI for comparison.

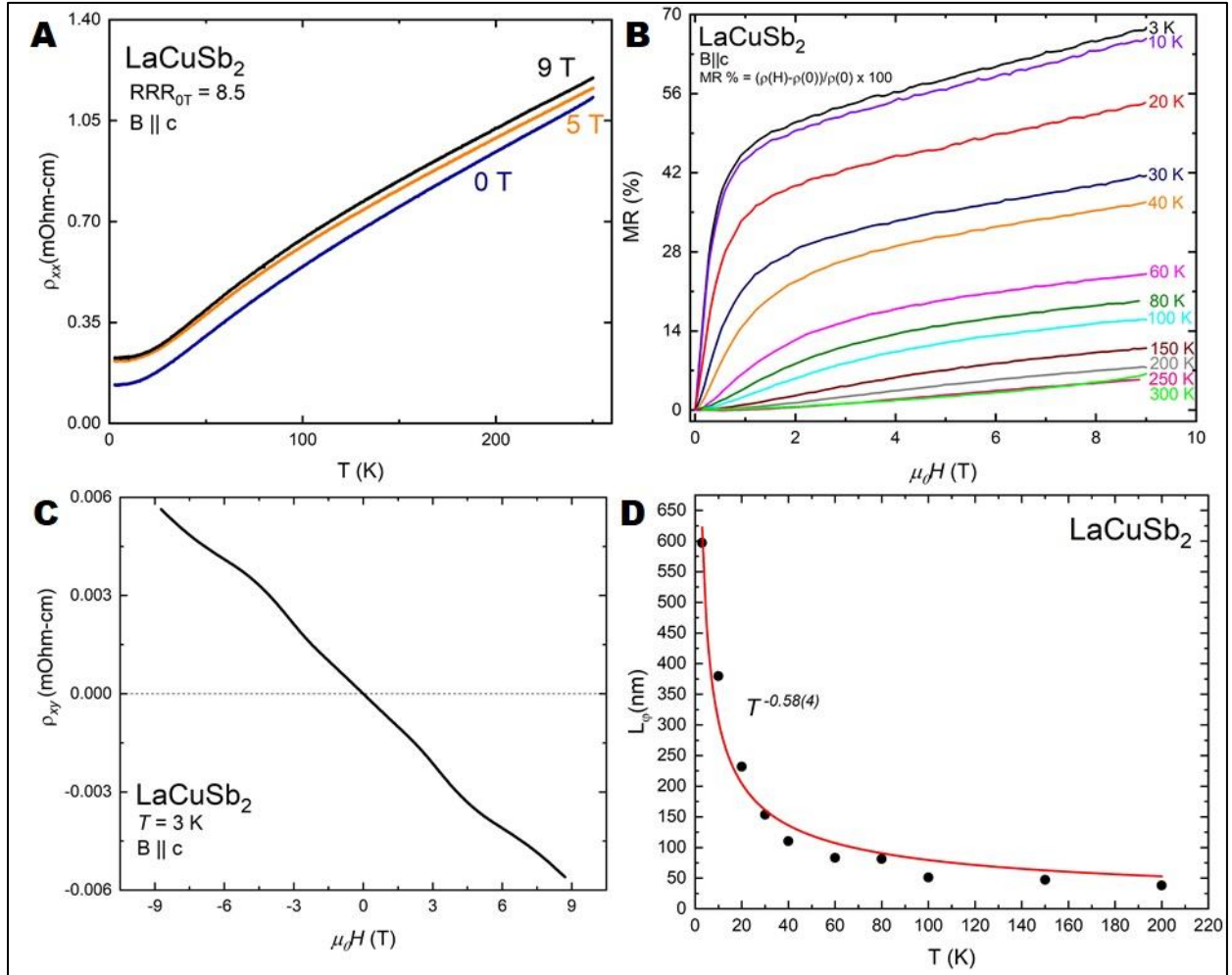

**Figure 3. A**. The resistivity as a function of temperature indicates metallic behavior, and generally increases upon applying magnetic fields at all temperatures. The RRR of 8.5 indicates the sample is of high quality. **B**. The magnetoresistance of LaCuSb$_2$ increases dramatically up to around ~2 T at the lowest measured temperature of $T = 3$ K, after which the dependence becomes linear with applied field. The region of linearity decreases with increasing temperature, but is somewhat regained at all fields past 200 K. The dramatic increase in magnetoresistance at low fields and low temperatures is due to the presence of weak antilocalization in this system. **C**. Measurements of the Hall effect in LaCuSb$_2$ suggest electrons to be the main carriers in this system. The non-linearity of the curve is due to issues with data collection and symmetrization. **D**. The phase coherence length of carriers in LaCuSb$_2$, as extracted by fitting the low-field magnetoresistance curves of LaCuSb$_2$ to the HLN theory for weak antilocalization. The $T^{-0.58(4)}$ power law dependence of $L_\varphi$ with temperature indicates that the weak antilocalization is two-dimensional.

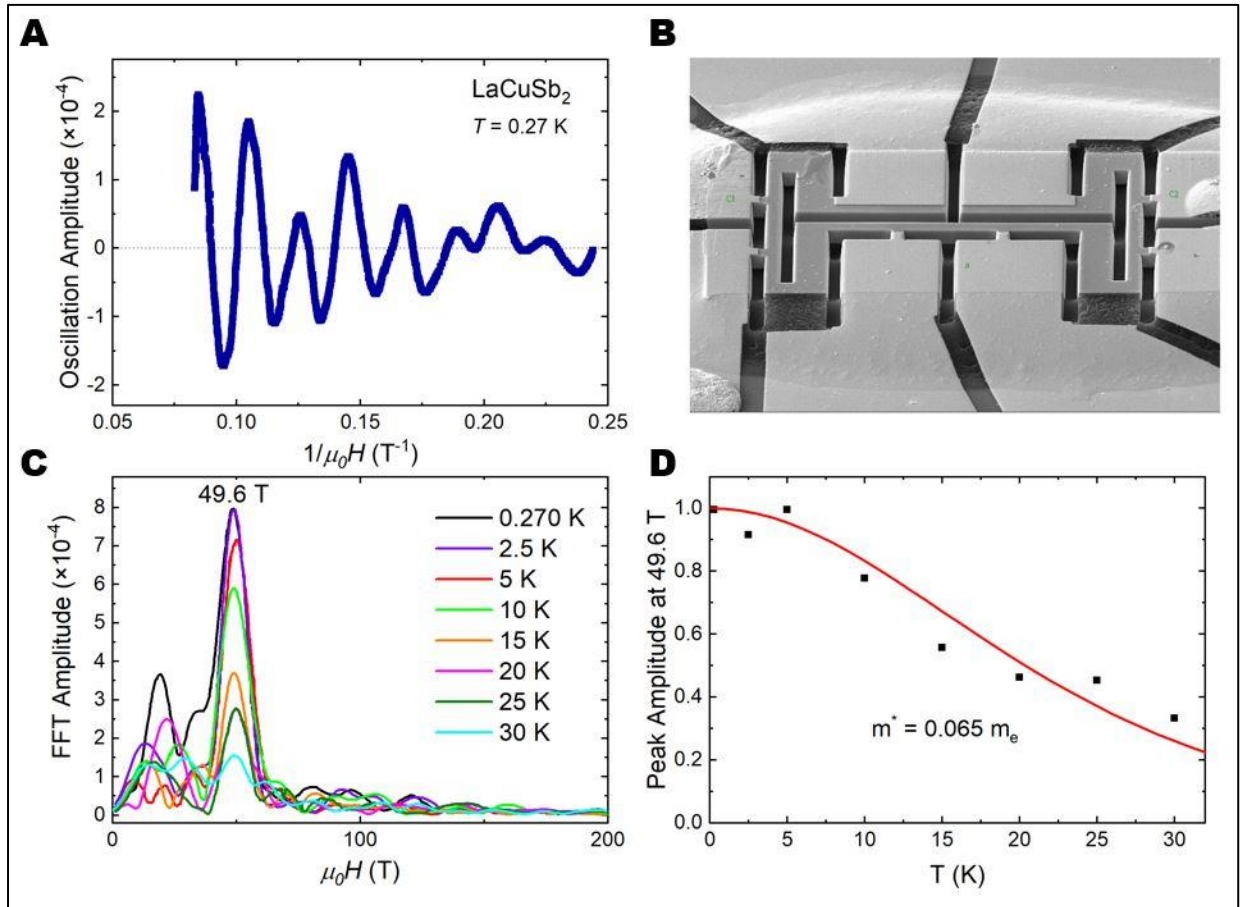

**Figure 4**. **A**. Oscillation amplitude as a function of inverse field at $T = 0.27$ K for LaCuSb$_2$ demonstrates large oscillations at high fields. **B**. A scanning electron microscopic image of the focused-ion beam cut device of LaCuSb$_2$ used for SdH oscillations measurements. **C.** Plot showing the Fast Fourier Transform (FFT) of the SdH oscillations at eight different temperatures. While several frequency amplitudes are observed, a frequency peak at 49.6 T is observed at all temperatures. **D.** The evolution of the 49.6 T frequency peak, shown in **C**, with temperature. A fit to Lifshitz-Kosevich theory yields an effective mass of 6.5% the electron rest mass.

J. R. Chamorro[1,2], A. Topp[3,4], Y. Fang[5], M. J. Winiarski[1,2,6], C. R. Ast[3], M. Krivenkov[7], A. Varykhalov[7] B. J. Ramshaw[5], L. M. Schoop[3,4], and T. M. McQueen[1,2,8,†]

[1]Department of Chemistry, The Johns Hopkins University, Baltimore, MD 21218, USA

[2]Institute for Quantum Matter, Department of Physics and Astronomy, The Johns Hopkins University, Baltimore, MD 21218, USA

[3]Max-Planck-Institut für Festkörperforschung, Heisenbergstraβe 1, D-70569 Stuttgart, Germany

[4]Department of Chemistry, Princeton University, Princeton, NJ 08544, USA

[5]Laboratory of Atomic and Solid State Physics, Cornell University, Ithaca, NY 14853, USA

[6]Faculty of Applied Physics and Mathematics, Gdansk University of Technology, ul. Narutowicza 11/12, 80-233 Gdansk, Poland

[7]Helmholtz-Zentrum Berlin für Materialien und Energie, Elektronenspeicherring BESSY II, Albert-Einstein-Straße 15, 12489 Berlin, Germany

[8]Department of Materials Science and Engineering, The Johns Hopkins University, Baltimore, MD 21218, USA

[†]mcqueen@jhu.edu

**ARPES Interpretation**

ARPES measures only a narrow range of $k_z$ values for a given photon energy. **Figure S1**, therefore, aims to determine the influence of a possible $k_z$ dispersion on the measured data. **Figure S1a** shows the experimental dispersion along the path described in the main text, and **Figure S1b** and **S1c** show the DFT band structures, with SOC included, along the corresponding paths for $k_z = 0$ and $k_z = \pi$, respectively. In the experimental data, intensity ranges were chosen independently for the path segments to ensure maximum visibility of the band structure. Along $\overline{\Gamma}$-$\overline{X}$, several parallel bands dispersing to lower initial state energies can be observed. For better visibility, a pair of arrows indicates one such pair of bands which is crossing the Fermi level, but shows rather weak intensity. These bands have no direct correspondence along the Z-R direction, but are clearly represented in the Γ-X DFT data of **Figure S1b**, certifying our choice of the $k_z = 0$ plane in the main paper. The X-M direction also catches the dispersion of the experimental bands along $\overline{X}$-$\overline{M}$, where they disperse parabolically towards $\overline{M}$. However, one must note that the energy gap between the first and second pair of bands along X-M is slightly too large to match the ARPES data perfectly. Since the ARPES data, for a photon energy of 80 eV, were not recorded perfectly on the high-symmetry plane $k_z = 0$, this deviation can be explained by the $k_z$ dispersion. The R-A direction in **Figure S1c**, for example, matches the small energy gap much better. This does not mean that the ARPES data follows the $k_z = \pi$ plane here, but instead suggests that its influence is strong in this part of the band structure when slightly off the BZ center.

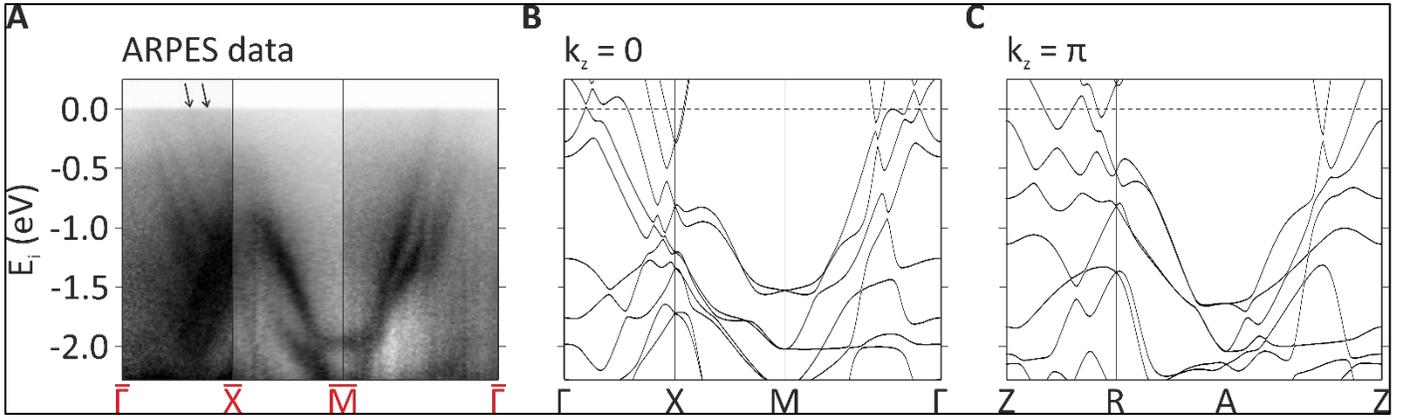

**Figure S1. A.** The experimentally observed band structure for the $\bar{\Gamma}$-$\bar{X}$-$\bar{M}$-$\bar{\Gamma}$ path along the Brillouin zone. **B.** The calculated band structure for the $k_z = 0$ plane. **C.** The calculated band structure for the $k_z = \pi$ plane. While the $k_z = 0$ case matches the data well, there are some contributions of the $k_z = \pi$ plane close to the $\bar{M}$ point.

**Two-Band Model Fit to Magnetoresistance**

The two-band model can be used to describe the magnetoresistance of a material with significant contributions from two different bands [1], and offers a possible alternative interpretation of the magnetoresistance of LaCuSb$_2$. The transverse resistivity $\rho_{xx}$ can be described by:

$$\rho(H) = \frac{(\sigma_{b1} + \sigma_{b2}) + \sigma_{b1}\sigma_{b2}(\sigma_{b1}R_{b1}^2 + \sigma_{b2}R_{b2}^2)H^2}{(\sigma_{b1} + \sigma_{b2})^2 + \sigma_{b1}^2\sigma_{b2}^2(R_{b1} + R_{b2})^2 H^2} = \rho_0 + \frac{\alpha H^2}{1 + \beta H^2}$$

Where $\rho_0$ is the zero-field resistivity, $\sigma_{b1}(\sigma_{b2})$ is the conductivity of carriers that occupy band 1 (band 2), $R_{b1}(R_{b2})$ is the Hall coefficient for band 1 (band 2) carriers, and $\alpha$ and $\beta$ are parameters defined as:

$$\alpha = \frac{(\sigma_{b1} + \sigma_{b2})\sigma_{b1}\sigma_{b2}(\sigma_{b1}R_{b1}^2 + \sigma_{b2}R_{b2}^2) - \sigma_{b1}^2\sigma_{b2}^2(R_{b1} + R_{b2})^2}{(\sigma_{b1} + \sigma_{b2})^3}$$

$$\beta = \frac{\sigma_{b1}^2\sigma_{b2}^2(R_{b1} + R_{b2})^2}{(\sigma_{b1} + \sigma_{b2})^2}$$

A fit of this model to the magnetoresistance data of LaCuSb$_2$ at $T = 3$ K is shown below:

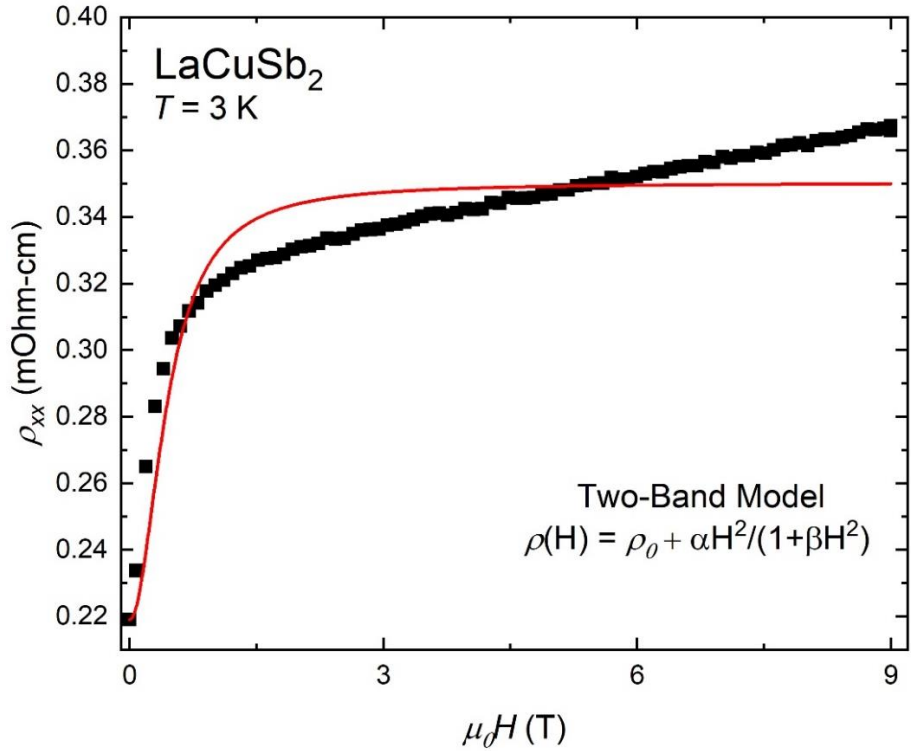

The simple two-band model does not account for the linearity in the magnetoresistance at fields above 2 T. Explicitly including a linear term in the model, however, improves the quality of the fit to the data, as shown below.

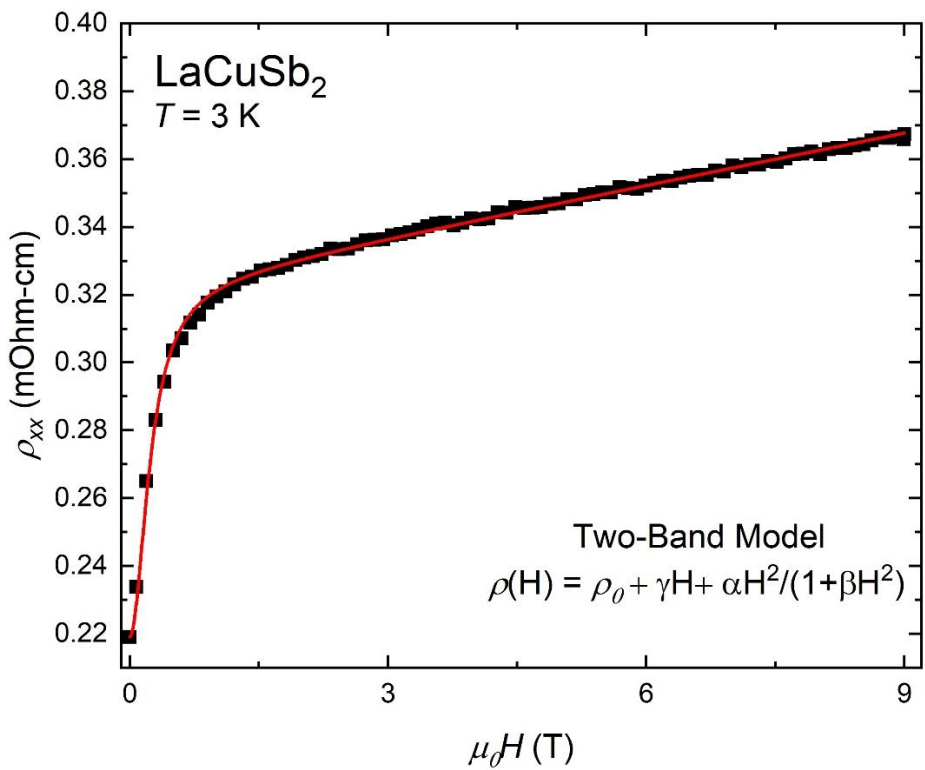

We therefore cannot rule out a two-band model explanation for the magnetoresistance if a linear term, such as is known to arise from Dirac Fermions or density fluctuations [2], is explicitly included.

**References**

[1] N. W. Ashcroft and N. D. Mermin, *Solid State Physics*, Saunders PA (1976).

[2] T. Kouri, *et. al.*, *Phys. Rev. Lett.* **117**, 25601 (2016).